\documentclass[prd,aps,twocolumn,a4paper,showkeys,nofootinbib,showpacs]{revtex4-1}

\usepackage{subfigure}

\usepackage{graphicx,psfrag}
\usepackage{mathrsfs}
\usepackage{amsmath,amsfonts,amssymb}
\usepackage{multirow}
\usepackage{comment}

\usepackage{xspace}
\usepackage{hyperref}
\usepackage{enumitem}

\newcommand{\be}{\begin{equation}}
\newcommand{\ee}{\end{equation}}
\newcommand{\bea}{\begin{eqnarray}}
\newcommand{\eea}{\end{eqnarray}}
\newcommand{\bel}{\begin{align}}
\newcommand{\eel}{\end{align}}

\newcommand{\gra}{\texttt{GR-Athena++}}

\newcommand{\tBAM}{\texttt{BAM}}
\newcommand{\BAM}{\tBAM\xspace}

\newcommand{\athenak}{\texttt{AthenaK}}

\def\p{\partial}

\def\gccm{{\rm g\,cm^{-3}}}

\def\GMc2{{\rm G M_{\odot} c^{-2}}}

\def\kt2{\kappa^\text{T}_2}

\def\Mo{{\rm M_{\odot}}}

\def\tD{\widetilde{D}}

\def\2nd{2^\mathrm{nd}}
\def\4th{4^\mathrm{th}}
\def\6th{6^\mathrm{th}}
\def\8th{8^\mathrm{th}}

\def\z4c{$\mathrm{Z}4\mathrm{c}$}
\def\z4oc{$\mathrm{Z}4(\mathrm{c})$}
\def\z4{$\mathrm{Z}4$}
\def\ccz4{$\mathrm{CCZ}4$}

\def\m1{$\mathrm{M}1$}

\usepackage{color}
\definecolor{cyan}{rgb}{0,0.9,0.9}
\definecolor{orange}{rgb}{0.9,0.5,0}
\definecolor{purple}{rgb}{0.8,0.4,0.8}
\definecolor{grey}{rgb}{0.8242,0.8242,0.8242}
\definecolor{brickred}{rgb}{0.8, 0.25, 0.33}
\definecolor{magenta}{rgb}{1,0,1}

\begin{document}

\title{Exact Mass Conservation in Binary Neutron Star Merger Simulations}

\author{Boris \surname{Daszuta}$^{1}$}
\author{Sebastiano \surname{Bernuzzi}$^{1}$}
\author{Joan \surname{Fontbuté}$^1$}
\author{Ruocheng \surname{Zhai}$^{2,4}$}
\author{Alan Tsz-Lok \surname{Lam}${}^{2,3}$}
\author{Jacob \surname{Fields}$^{5}$}
\author{David \surname{Radice}${}^{2,3,4}$}
\affiliation{${}^1$Theoretisch-Physikalisches Institut, Friedrich-Schiller-Universit{\"a}t Jena, 07743, Jena, Germany}
\affiliation{${}^2$Institute for Gravitation and the Cosmos, The Pennsylvania State University, University Park PA 16802, USA}
\affiliation{${}^3$Department of Physics, The Pennsylvania State University, University Park PA 16802, USA}
\affiliation{${}^4$Department of Astronomy \& Astrophysics, The Pennsylvania State University, University Park PA 16802, USA}
\affiliation{${}^5$Institute for Advanced Study, Princeton, NJ 08540, USA}

\date{\today}

\begin{abstract}
  A long-standing problem in the simulation of neutron star spacetimes
  is the treatment of vacuum regions outside the stars. The use of an
  artificial low-density atmosphere is a common robust approach within
  Eulerian hydrodynamics that, however, introduces baryon-mass violation
  even with conservative numerical schemes. We propose a simple
  numerical algorithm that ensures exact mass conservation by means of an
  appropriate local rescaling of the atmosphere. The scheme is
  combined with a low-order flux correction
  and it can be further augmented by a pseudo-vacuum treatment that
  enforces strict vacuum in the outer regions far from the central objects.
  We demonstrate the effectiveness of these vacuum treatments with
  binary neutron star merger simulations spanning multiple orbits and
  the postmerger phase, and including a microphysical equation of state.
  The rescaling algorithm guarantees mass and electron number
  conservation to round-off precision. The pseudo-vacuum treatment
  shows slightly larger but approximately constant violations and can
  improve the computation of fast tail ejecta 
  as well as provide convergent gravitational waves of quality comparable to the 
  standard atmosphere.
  Overall, results from different atmosphere treatments and a two-code comparison 
  suggest that current computations of gravitational waves and
  (dynamical) ejecta in the presence of an artificial atmosphere are robust,
  provided that conservative adaptive mesh refinement with 
  flux correction is employed.
\end{abstract}

\maketitle

\section{Introduction}
\label{sec:intro}

The simulation of neutron star spacetimes requires handling vacuum
regions and the dynamical transition between vacuum and fluid on the computational
domain, see e.g.~discussions in \cite{Font:2001ew,Dimmelmeier:2002bk,Baiotti:2004wn,Thierfelder:2011yi,Cook:2023bag}.
For example, the exterior spacetime of a compact star is
vacuum and needs to be included in the computational domain.
Vacuum encompasses the spatial regions distant from the strong-field regime
of a binary neutron star merger (BNSM), where gravitational waves (GWs)
are extracted.
Matter ejecta unbound during the merger dynamics 
expand and fill these vacuum regions along the evolution.
Vacuum regions can also appear after black hole formation as matter
flows in and accretes onto the black hole. 
The treatment of vacuum regions is a central practical issue to obtain
successful numerical-relativity simulations of astrophysical
interest.

The most common approach to neutron star simulations is conservative Eulerian
hydrodynamics with high-resolution shock capturing schemes. In this context, it is
standard practice to substitute vacuum with a dilute atmosphere and let
the numerical scheme handle sharp gradients at the
matter-vacuum interface.
While robust solutions are available, this procedure introduces
artifacts. Baryon (rest) mass violation is commonly observed even when
finite volume (FV) schemes and conservative adaptive mesh refinement
(AMR) techniques are employed. 
This violation may, in principle, impact the accurate computation of
mass ejecta and gravitational waves and introduce a slowly (if at all)
converging systematic error in the computations. 
Similarly, sound waves, velocity gradients up to ${\sim}0.6$c and
temperature gradients up to a few MeV can develop at matter-vacuum
interfaces in BNSM altering the thermodynamical state of the simulated matter,
e.g.~\cite{Kastaun:2020uxr,Hammond:2021vtv}, and potentially affecting
microphysics (ejecta composition, weak-reaction rates, etc).  

In Newtonian hydrodynamics the matter-vacuum transition can be
handled by specific Riemann solvers and/or level set
methods~\cite{Toro:1999,Leveque:2002}. However, this approach significantly
complicates the algorithms, especially with nontrivial grid geometries
or AMR. This is even more serious in presence of gravity, where the
treatment of vacuum regions can alter hydrostatic equilibria between the fluid's pressure 
and the gravity pull. Specific well-balanced numerical schemes have
been introduced to exactly preserve hydrostatic equilibrium solutions,
and were found essential to simulate astrophysical fluids,
e.g.~\cite{Chandrashekar:2015a,Berberich:2020a,Edelmann:2021a} and
references therein.

In general-relativistic hydrodynamics, numerical issues are
amplified by the errors introduced during the numerical inversion of
conservative to primitive variables. Such transformation is (typically) not
analytical and failures during the inversion (even floating-point drift) handled via flooring can cause violation of mass conservation. 
Non-conservative AMR treatment is another cause of mass
conservation violation~\cite{East:2011aa,Reisswig:2012nc,Dietrich:2015iva}. 

While these are very well known problems for practitioners, no
rigorous and robust alternative exists so far and the use of an 
artificial atmosphere is the most common approach in neutron stars
simulations.
A vacuum treatment has been proposed in 
\citet{Poudel:2020fte} (see also~\cite{Tichy:2022hpa}) 
indicating some advantages for neutron star simulations with the \BAM{}
code~\cite{Brugmann:2008zz,Thierfelder:2011yi}. However, the method
still violates mass conservation.  
More recently, the first-order flux correction (FOFC) approach of
\citet{Lemaster:2008gh} has been applied to BNSM simulations with 
\athenak{}~\cite{Stone:2024,Fields:2024pob}.
The method significantly reduced mass violation and appears to improve
simulation robustness.

In this paper, we explore a new numerical algorithm to ensure exact mass
conservation in neutron star spacetime simulations.
The scheme relies on an appropriate rescaling of the standard
atmosphere treatment and can be easily incorporated in existing codes.
It can be further augmented by a pseudo-vacuum treatment
that enforces strict vacuum in the outer regions far from the central objects.
These methods are implemented in the \gra{} AMR 
code together with a low-order flux
correction~\cite{Daszuta:2024chz,Cook:2023bag,Daszuta:2024ucu}.  

The paper is organized as follows. Section~\ref{sec:ome} introduces
the details of the rescaling algorithm, of the low-order flux
correction and the pseudo-vacuum treatment. 
Section~\ref{sec:application} presents first results for BNSM
simulations with a focus on mass
ejecta and GWs. A direct comparison with \athenak{} is also discussed.
Conclusions are in Sec.~\ref{sec:conclusion}

\section{Method}
\label{sec:ome}

We focus on the (3+1)D balance-law formulation of Eulerian general-relativistic
hydrodynamics (GRHD)~\cite{Banyuls:1997zz}. The local conservation law for
the Eulerian mass-density~\footnote{Tilde always
indicates densitization by the spatial 3-metric  
$\gamma_{ij}$; i.e. $\tD =\sqrt{\det\gamma}D$. Einstein's index-summation convention
is understood with $i=1,2,3$.},  
\be\label{eq:D}
\p_t\tD + \p_i F^{(i)}_{D} = 0
\ee
where $F_{\tD}$ is the appropriate flux, guarantees baryon
(rest) mass conservation,
\be
M_b = \int_{\Sigma_t} \mathrm{d}^3x \tD, \qquad \dot{M}_b(t) = 0, 
\ee
if the integral is taken on the entire 3D spatial hypersurface
$\Sigma_t$. When the fluid composition is incorporated in the equation
of state (EOS), as common, hydrodynamics equations are augmented by
conservation or balance laws for particle species. For example, the
number of electrons per baryons (nucleons) $Q_e$ is conserved
according to the local conservation law for the electron fraction $Y_e$,
\be\label{eq:Ye}
\p_t(\tD Y_e) + \p_i F^{(i)}_{Y_e} = 0.
\ee
If neutrino radiation is simulated (outside the scope of this paper), $Y_e$ evolves
based on weak-interaction rates introduced as source terms at 
the r.h.s. of Eq.~\eqref{eq:Ye}. 

Conservative FV methods are designed precisely to preserve these 
continuum properties at the discrete level.
Nonetheless during numerical computations, the values of
$M_b$ and $Q_{\mathrm{e}}$ can (and usually do) drift beyond what can be ascribed to outflow through the computational domain.
The origin of the violation is usually the treatment of vacuum region
and the use of non-conservative AMR.
Note that violations are not limited to floating-point errors but are
often due to failures during primitive reconstruction, in particular if the
variable inversion is handled with flooring \cite{Kastaun:2020uxr,Cook:2023bag}.

\subsection{Standard atmosphere}
\label{sbsbsec:standard_ap}

A standard approach to deal with vacuum regions is to adjust the
initial data through the introduction of an unphysical, static
atmosphere, over each cell where the matter density is zero. 
The atmosphere points are then evolved in the same way as the other fluid
cells. Atmosphere floors may be applied at various steps of the
evolution in order to avoid division by zero and prevent the
evolution towards unphysical states.
A common choice is to enforce atmosphere floors during the
conservative to primitive variable inversion. Here, a fluid cell that
has evolved below atmosphere level is reset to a floor value. In
case the numerical inversion fails, a point may also be reset to the
atmosphere floors. 

The choice of floor values is implementation dependent and
based on experimentation with hydrostatic problems, e.g. spherical neutron stars
spacetimes. 
The specific implementation adopted in \gra{} and considered for our
work is described in \cite{Kastaun:2020uxr,Cook:2023bag} and available open source~\footnote{
\url{https://computationalrelativity.github.io/grathenacode/}}.
For a composition-dependent equation of state (EOS), usually supplied
in tabular form, the atmosphere at a point is fixed  by imposing mass-density and
temperature floors, and setting $Y{}_{\mathrm{e}}$ to a prescribed (atmosphere) value,
$$
(D{}_{\mathrm{fl}},\,T{}_{\mathrm{fl}},\,Y{}_{\mathrm{e},\,\mathrm{atm}}).
$$
The choice of density and temperature floors is based on the minimum
tabulated value. In atmosphere, zero fluid velocity is assumed and all
the other quantities are computed consistently with the EOS.

\subsection{Rescaling}
\label{sbsbsec:rescaling}

As a remedy to drifts in $M_b$ and $Q_e$ we propose a
simple rescaling procedure of the atmosphere that keeps under control
the violation in conserved quantities.
At any timestep we compute the violation in conservation
based on the difference between current and initial value,
$\delta M_b(t) =M_b(t) - M_b(0)$ (and similarly for $Q_e$.)
After a timestep, the violation is then driven to zero, $\delta M_b\to
0$, by rescaling the atmosphere over regions containing matter at
near-atmosphere levels. To achieve this we proceed with an iterative
prescription.

We define the ``cutoff integral''
\be
\mathcal{I}[D;\,f_D]:=\int_{D < f_D
  \min(\tD)}\,\mathrm{d}^3x\,\tD,
\ee
where the parameter $f_D>1$ serves the role of controlling the upper
bound for data that will be rescaled.
For a chosen cutoff $f_D$ we iteratively set $f_D\leftarrow f_D^2$
until $\mathcal{I}(D;\,f_D)>\delta M_b$ is satisfied~\footnote{%
We evaluate this via a numerical quadrature and employ
Kahan-Babushka-Neumaier compensated summation
\cite{Neumaier:1974}.}.
Finally, at all points such that $D\leq f_D \min(D)$ the
mass density is rescaled by 
\be\label{eq:src:algo}
\tD\leftarrow \tD\times \frac{M_b(0) - \mathcal{I}[D;\,f_D]}{M_b(t) - \mathcal{I}[D;\,f_D]}.
\ee
The overall procedure is terminated if the mass (or $Q_e$) violation 
deviates excessively from its initially recorded value. This is
determined by monitoring the absolute relative error of the
corresponding conserved quantity with respect to its initial value. A
threshold of $10^{-4}$ was chosen to disable rescaling for all salient
runs.
We stress that the rescaling assumes no outflow from the computational domain
and absence of sources or sinks in the domain of integration.  

The rescaling we have described may be combined with standard
atmosphere treatments. It may also be used in conjunction with the
approach described in Sec.~\ref{ssec:layer} which we term
pseudo-vacuum. 

\subsection{Pseudo-vacuum}
\label{ssec:layer}

One alternative to global imposition of atmosphere is to only
introduce floors in regions local to matter. Prior to evaluation of a 
time-integrator stage, cells with $D>0$ are flagged as ``matter
containing''. Floors are then only imposed on the remaining cells if any
nearest-neighbour~\footnote{Our implementation leaves this freely
specified; here we only consider immediate nearest-neighbours.} cell
contains matter. This introduces a thin layer of atmosphere. The time
sub-step is then calculated for the flagged cells, with the layer
providing support for flux computations. After a full time-step, any
cell whose density falls below a prescribed cutoff $D_{\mathrm{cut}}$
is reset to strict vacuum; no rescaling is imposed during sub-steps.

Applying the above procedure, allows matter to flow into regions
initially without matter, with spatial extent developing dynamically
according to the simulation. In vacuum regions, the matter field
evolution is thereby entirely trivialized, which also serves to reduce
computational cost of the matter sector evaluation in proportion to
the relative size of the inactive region. A similar strategy is implemented 
in the \BAM{} code, although we note that \BAM{} AMR is not exactly conservative~\cite{Thierfelder:2011yi,Poudel:2020fte,Dietrich:2015iva}. 

\subsection{AMR implementation with Low-order Flux correction}
\label{sbsbsec:lofc}

The algorithms described above have been implemented in
the \gra{} code~\cite{Daszuta:2024chz,Cook:2023bag,Daszuta:2024ucu}.
\gra{} employs high-resolution shock-capturing (HRSC) schemes based on conservative
FV methods and block-based AMR. The code has been extended with a
low-order flux-correction (LOFC), further supplemented by an approximate
discrete-maximum-principle (DMP)~\cite{Daszuta:2026szb}, as motivated by improved robustness and
conservation properties observed with the first-order flux correction
(FOFC) in \cite{Stone:2024,Fields:2024pob}. 
We briefly summarize the procedure below.

At each time sub-step, numerical fluxes at interfaces are assembled
based on (primitive) reconstruction (here WENO5Z \cite{Borges:2008a}),
and combined according to the local-Lax-Friedrichs (LLF) prescription
\cite{Zanna:2002qr}.
This gives rise to a candidate state vector for the conserved Eulerian
variables 
(prior to addition of any geometric source-terms).
If the candidate state does not satisfy DMP, or would require
atmosphere flooring
(Sec.~\ref{sbsbsec:standard_ap}), the cell is flagged for LOFC. If
flagged, for all interfaces adjacent to the problem-cell, we switch to
low order primitive reconstruction (here PLM-van Leer
\cite{vanLeer:1974a}), and recompute the
LLF fluxes. This flux correction remains conservative, and is used to
propagate the state instead. The LOFC is applied to all simulations
presented below.

\section{Application to BNSM}
\label{sec:application}

\begin{figure*}[t]
  \centering
  \includegraphics[width=\textwidth]{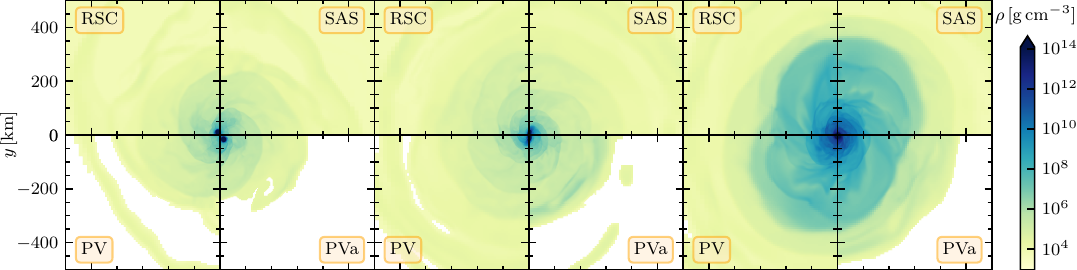}
  \includegraphics[width=\textwidth]{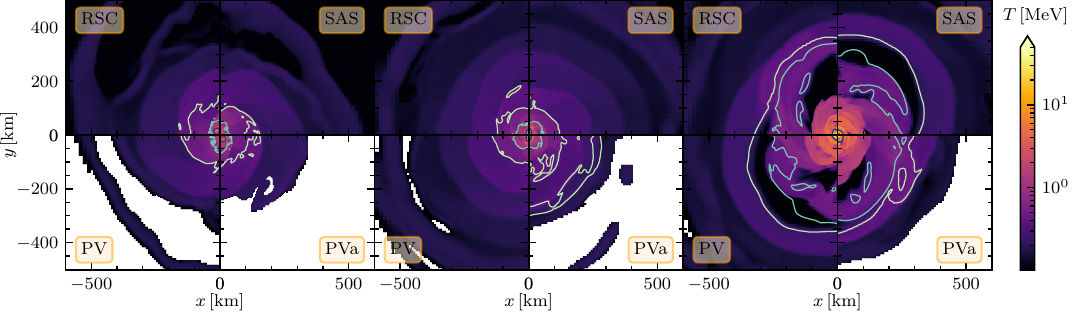}
  \caption{Rest-mass density $\rho$ (top) and temperature $T$ (bottom)
    evolution in the orbital plane for four atmosphere treatments. 
    Each column shows a snapshot of the remnant at fixed time where
    column-wise we have: $t-t_{\rm
      mrgr}\simeq(-4.8,\,0.4,\,4.2)\,\mathrm{ms}$. At fixed time, four
    quadrants are displayed for $\rho$ and $T$. For each variable we
    show in the (upper right quadrant) the standard atmosphere
    treatment (SAS), (upper left) rescaling, (lower left)
    pseudo-vacuum (PV) with $D_{\mathrm{cut}}=9.7\times10^3\gccm$, and
    (lower right) PV with more aggressive $D_{\mathrm{cut}}=32\times
    10^3\gccm$ denoted (PVa). For the $T$ panels we additionally
    overlay isodensity contours at:
    $\rho=10^{12}\,\mathrm{g}\,\mathrm{cm}^{-3}$ (dark blue),
    $\rho=10^{7}\,\mathrm{g}\,\mathrm{cm}^{-3}$ (pale green), and
    $\rho=5\times10^{5}\,\mathrm{g}\,\mathrm{cm}^{-3}$ (pale yellow). 
    Data depicted here correspond to simulations with grid-setup
    featuring finest level resolution of $\delta x_f$.}
 \label{fig:rho_tmp_xy}
\end{figure*}

We now compare the standard, rescaling, and pseudo-vacuum (PV)
atmosphere treatments discussed in Sec.~\ref{sec:ome} in BNSM simulations.
To this end, we consider the evolution of equal-mass quasicircular irrotational
BNS data with mass $M=2.7~\Mo$; initial data
are prepared based on the \texttt{Lorene} solver
\cite{Gourgoulhon:2000nn}.
Neutron star matter is described by the
temperature-dependent DD2 EOS \cite{Hempel:2011mk}.
The DD2 binary evolves for ${\sim} 4.5$ orbits and produces a long-lived
remnant; the simulations are further conducted until at least
$20~\mathrm{ms}$ postmerger. 

The computational domain is a Cartesian mesh over the volume defined
by $[-2268,\,2268]^3~\mathrm{km}$ with AMR setup that closely
follows the cell-centered setup of \citet{Daszuta:2024ucu}.
The domain is sampled with $N_M=96$ or $144$ points in each
dimension, decomposed in mesh blocks of fixed sampling $16^3$
points and recursively refined (at $2:1$
refinement-ratio) until a target, minimum resolution is achieved.
The finest resolutions considered for three setups are $\delta
x_c\simeq492.2\,\mbox{m}$, $\delta x_m\simeq369.2\,\mbox{m}$,
and $\delta x_f\simeq246.1\,\mbox{m}$. 
No mesh symmetries are imposed.
Time stepping is performed with strong-stability preserving Runge-Kutta SSPRK(3,3)~\cite{Gottlieb:2009a}
and Courant-Friedrich-Lewy
condition of $0.25$ is set.

We next discuss results for simulations performed at the three common resolutions for four distinct atmosphere treatments:  (i) the standard
atmosphere treatment (denoted SAS),
(ii) the rescaling algorithm with $f_D=10=f_{Y_\mathrm{e}}$ (denoted
RSC), (iii) the pseudo-vacuum algorithm with $f_D$ and
$f_{Y_{\mathrm{e}}}$ as in RSC, and cut chosen as
$D_{\mathrm{cut}}=9.7\times10^3\gccm$ (denoted PV), and (iv) the PV
algorithm with a more aggressive choice of $D_{\mathrm{cut}}=32\times 10^3\gccm$ (denoted PVa). For all simulations, floor and atmosphere parameters are set as
$(D{}_{\mathrm{fl}},\,T{}_{\mathrm{fl}},\,Y{}_{\mathrm{e},\,\mathrm{atm}})=
(6\times10^3\,\gccm,\,0.1\,{\rm MeV},\,0.5)$. We emphasize that each of these four setups is otherwise identical to
the others, only differing in the details of the atmosphere treatment. 

\subsection{Mass and Electron number conservation}
\label{sbsbsec:mass}

\begin{figure*}[t]
  \centering
  \includegraphics[width=0.7\textwidth]{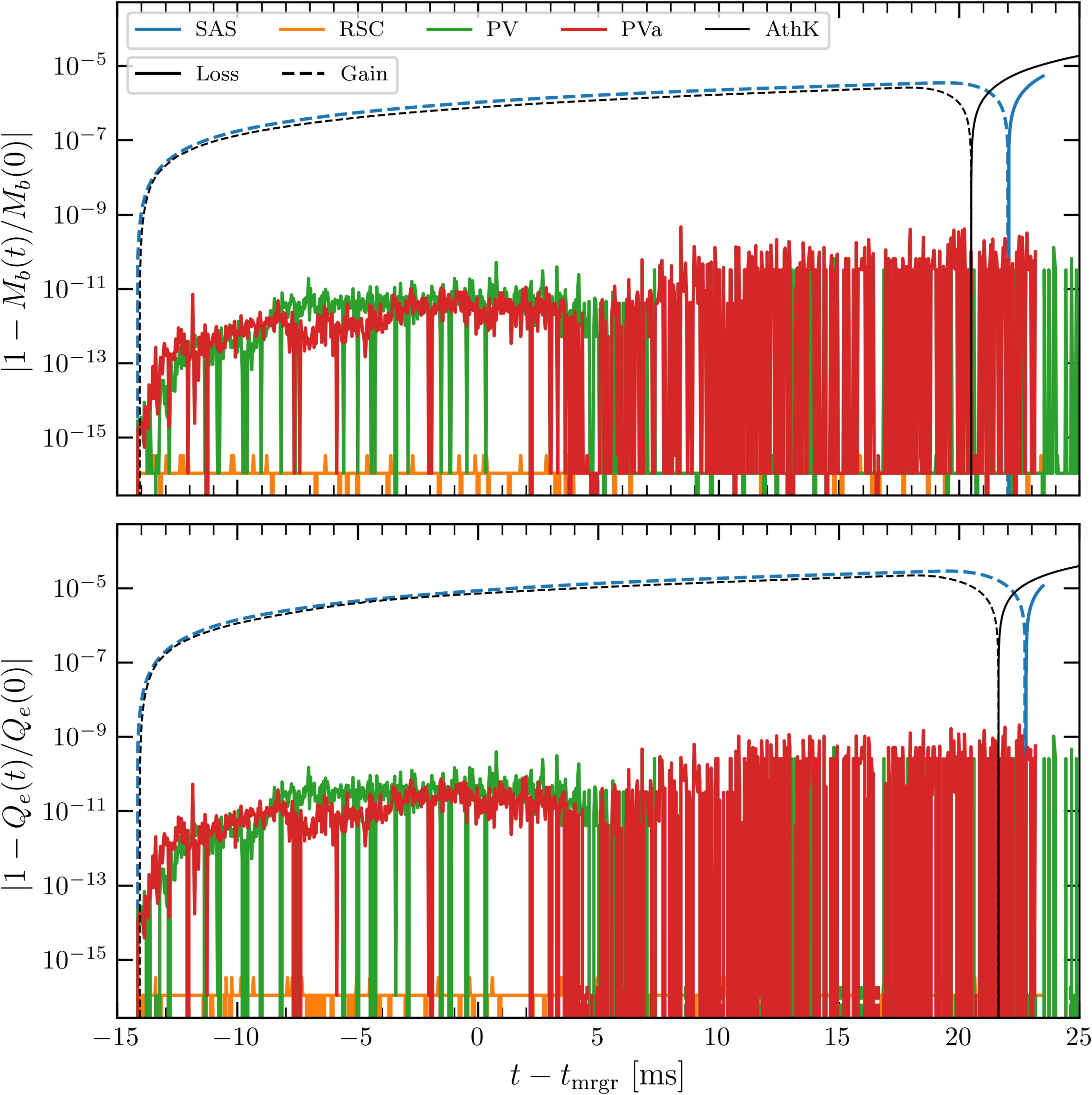}
  \caption{Evolution of the relative error in baryon (rest) mass and
    electron number conservation for the four vacuum algorithms at
    fixed finest resolution $\delta x_f$. Dashed (solid) lines highlight mass
    or electron number gains (losses).    
    The plot demonstrates exact (to round off error) conservation for
    both quantities with the rescaling algorithm (RSC) introduced in
    our work, and an approximately constant violation
    ${\lesssim}10^{-11}$ with both variants of pseudo-vacuum (PV, PVa)
    algorithm. The standard atmosphere treatment (SAS) introduces
    relative gains of order ${\sim}10^{-6}$ in both mass and electron
    number conservation. Simulations involving the lower resolutions
    $\delta x_c$ and $\delta x_m$ display comparable behaviour (not
    depicted here). \athenak{} data obtained with a very similar
    simulation setup as \gra{} SAS are shown in black.}
 \label{fig:cons}
\end{figure*}

Snapshots at distinct times of the evolution of the (fluid-frame)
density together with temperature in the orbital plane for the DD2
binary are shown in Fig.~\ref{fig:rho_tmp_xy} for the four atmosphere
treatments. As is common in all BNSM simulations, matter leaks  
out from the star surfaces before merger and fills a region around the stars at
densities \emph{above} the atmosphere threshold and temperatures of a few MeV.
The rescaling algorithm cannot avoid this effect, which is observed with all treatments.
At merger, matter compresses and bounces while the two cores fuse to
form a remnant with maximum densities $\rho_{\rm max}^{\rm
  remnant}\sim 1.3\rho_{{\rm max}}^{\rm NS}$, where $\rho_{\rm
  max}^{\rm NS}$ is the initial maximum density of the isolated NS.
Pressure waves propagating outwards
steepen into shock waves with densities ${\lesssim} \rho^{{\rm NS}}_{\rm
  max}$ and peak temperatures $T\sim70$~MeV~\cite{Perego:2019adq}.
These waves develop into spiral density
waves \cite{Nedora:2019jhl} and rapidly fill the entire computational
domain with matter densities above the atmosphere. 
These dynamics are observed for all atmosphere treatments with
negligible quantitative difference. Notably, even the PV treatments
do not alter the picture: the orbital motion and hydrodynamics flow
around remnants are all consistent at densities above atmosphere. 

Figure~\ref{fig:cons} shows the relative error in the baryon mass and
electron number conservation. The conservative AMR
and standard atmosphere treatment in \gra{} guarantees relative baryon mass
violations ${\lesssim}10^{-6}$ at the (moderate) resolution considered
for these simulations. The violation constitutes mass \emph{gain}. For
the runs presented here, with the standard atmosphere treatment, this can
be understood as follows.
In regions exterior to the binary constituents, prior to significant
generation of mass ejecta, matter is typically near-atmosphere. During 
the evolution, atmosphere accretes, thereby pushing $D$
below $D_{\mathrm{fl}}$. This in turn, leads to artificial injection
of matter as floors are applied. 
On the other hand, the rescaling algorithm delivers mass conservation
to round off precision, with exterior regions falling below the
dynamically inferred cut that enters the algorithm ($f_D$). 
The PV treatment shows similar mass conservation but to
${\lesssim}10^{-11}$ level. This larger violation compared to the
  rescaling is related to the mass violation during a time
  integrator timestep.

The conservation of electron number is also shown in
Fig.~\ref{fig:cons}, bottom panel. The behaviour closely mirrors the
behaviour of the mass, indicating that the largest violations in this
quantity are also due to the vacuum treatment. 

\begin{figure}[t]
  \centering
  \includegraphics[width=0.49\textwidth]{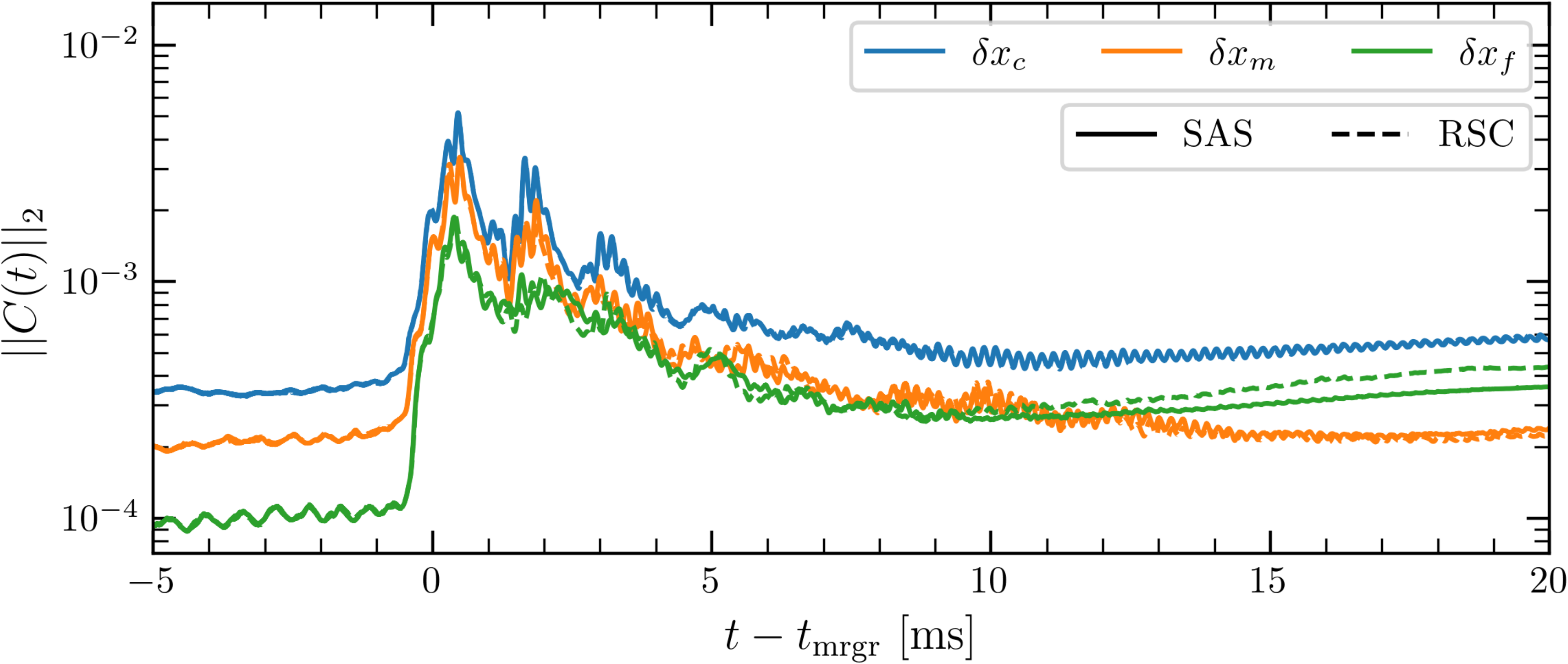} 
    \caption{Evolution of the $L_2$-norm of the Z4c collective
      constraint monitor $C(t)$ for DD2 BNS simulations, at a variety
      of resolutions, and different atmosphere treatments. The
      standard treatment (solid lines) and the rescaled (dashed lines)
      atmosphere treatments are seen to agree well, for runs with grids
      involving finest resolutions $\delta x_c$, $\delta x_m$, and
      $\delta x_f$.}
    \label{fig:constr}
\end{figure}

\gra{} employs the Z4c free evolution scheme for the spacetime
evolution~\cite{Bernuzzi:2009ex,Hilditch:2012fp}.
The numerical solution of free-evolution schemes introduces violation
of the Einstein constraints
as well as other constraints related to the specific hyperbolic formulation of Einstein
field equations. It is therefore important to assess that the new
vacuum treatments do not introduce further violations of such constraints.

Figure~\ref{fig:constr} shows the evolution of the $L_2$ norm of the
collective Z4c constraint monitor $C$ for the standard and rescaled atmosphere
treatment at the three mesh resolutions.
The specific expression for $C$ includes Hamiltonian
and momentum constraints as well as the four $Z$-constraints, see Eq.~(40)
of \cite{Weyhausen:2011cg} for its expression. Note the calculation of
the norm of the constraint monitor excludes points towards grid boundaries where
approximate Sommerfeld boundary can lead to relatively large local
violations (while being stable.)
The constraint monitor for the two treatments are consistent with each
other and converge to zero as resolution is increased.
The constraint violation in these simulations is largely dominated by
hydrodynamical truncation errors that are, in turn, dominated by the
bulk fluid motion (i.e. not by the constant atmosphere.)
The postmerger dynamics in particular is characterized by a more
complex flow and shocks and therefore typically shows slower convergence. 
Our novel treatment does not significantly impact on constraint violations.
Similar results are found for PV treatments, not shown in the figure.

\subsection{Ejecta}
\label{sbsbsec:ejecta}

\begin{figure}[t]
  \centering
  \includegraphics[width=0.49\textwidth]{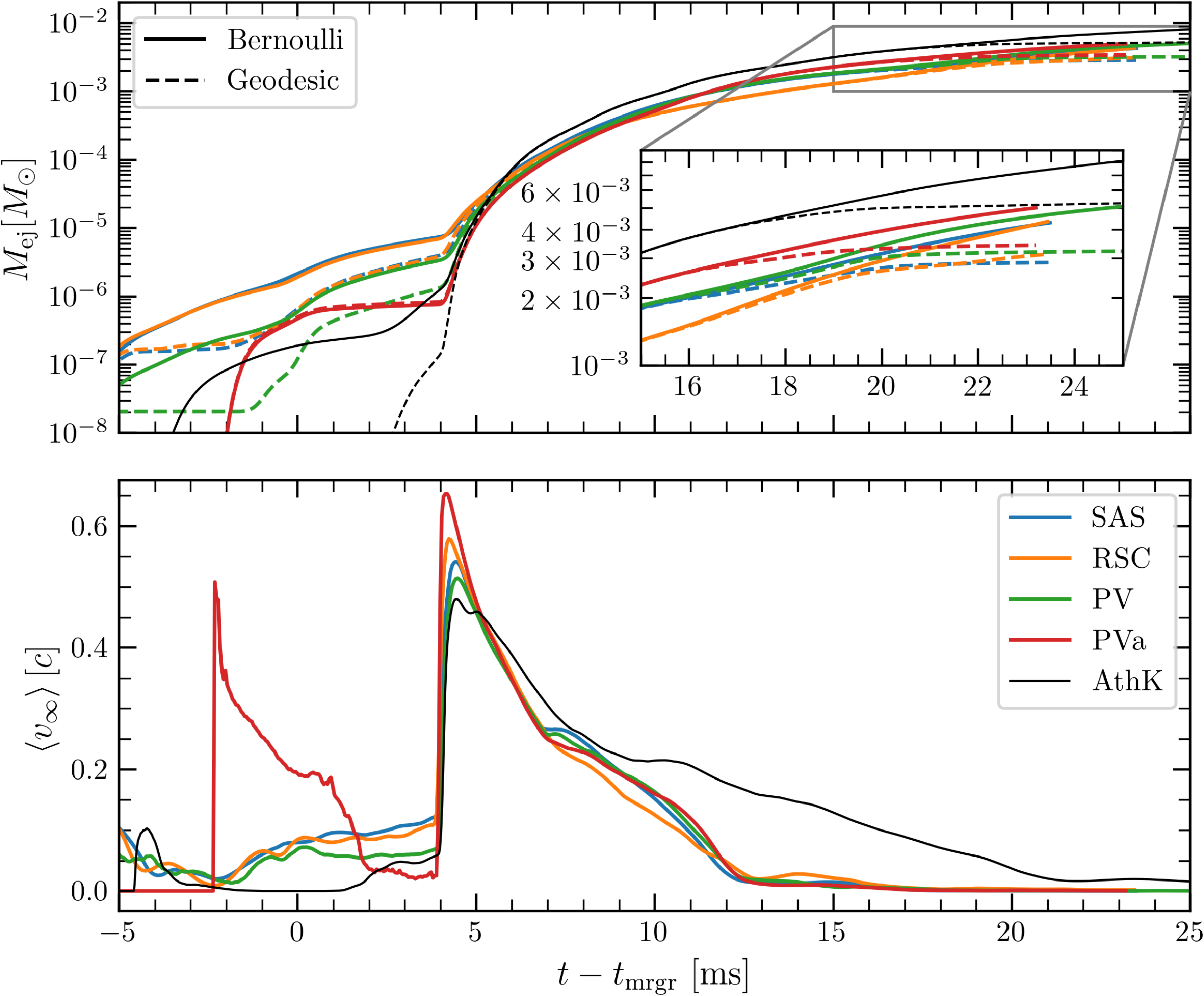}
  \caption{Evolution of ejecta mass (top) and mass-averaged velocity (bottom) for the four atmosphere algorithms.
    Unbound mass is monitored either according to Bernoulli criterion (solid
    lines) or geodesic criterion (dashed line) and extracted at a
    sphere of coordinate radius $R\simeq 591\,\mbox{km}$. Common
    (finest) resolution of these runs is $\delta x_f$. \athenak{} data
    obtained with a very similar 
    simulation setup as \gra{} SAS are shown in black.} 
  \label{fig:ejecta}
\end{figure}

During the first core bounce, part of the matter becomes unbound and
generates dynamical ejecta~\cite{Radice:2018pdn}. 
Figure~\ref{fig:ejecta} shows the evolution of the ejecta mass (top)
and the mass-averaged asymptotic velocity (bottom) for the lowest
resolution simulations. These quantities
are computed by flagging fluid elements that reach a sphere at
coordinate radius $R\simeq 591\,\mbox{km}$ with outwards-pointing velocity
using either the Bernoulli criterion,
i.e.~$hu_t < -h_{\mathrm{min}}$, where $h$ and $h_{\mathrm{min}}$ are
the specific and minimum (of the EOS table) enthalpies, and $u_t$ the 
time component of the fluid 4-velocity, or the geodesic criterion, $u_t < -1$.

At earlier times $t-t_{\rm mrgr}<5\,\mbox{ms}$ the mass ejection is spurious and mainly
related to low density material around the stars. Here, the Bernoulli
criterion significantly overestimates the ejecta,
cf.~\cite{Kastaun:2014fna}.
The standard and rescaled atmosphere treatments behave very similarly,
while the PV treatment reduces the early, spurious mass
ejection since no significant mass is present at the extraction sphere.
Note, however, the PV treatments can exhibit a larger velocity peak due to clumps of material
propagating in the pseudo vacuum.

The bulk of the dynamical ejecta reaches the coordinate sphere at
$t\sim5\,\mbox{ms}$ postmerger. At these times the mass rises quickly by about an
order of magnitude to
${\sim}10^{-4}\,\Mo$ and the averaged asymptotic velocity peaks at ${\sim}0.4-0.6$~c,
independently of the atmosphere treatment. The dynamical ejecta
quickly saturate afterwards to a total mass of
${\sim}5\times10^{-3}\,\Mo$. Their velocity rapidly decreases 
towards zero (more on this below).
At later postmerger times $t\gtrsim20\,\mbox{ms}$, the dynamical
ejecta terminates, and
all simulations show the beginning of the early spiral-wave wind
from the remnant. The latter is captured only by the Bernoulli
criterion.

\begin{figure}[t]
  \centering
  \includegraphics[width=0.49\textwidth]{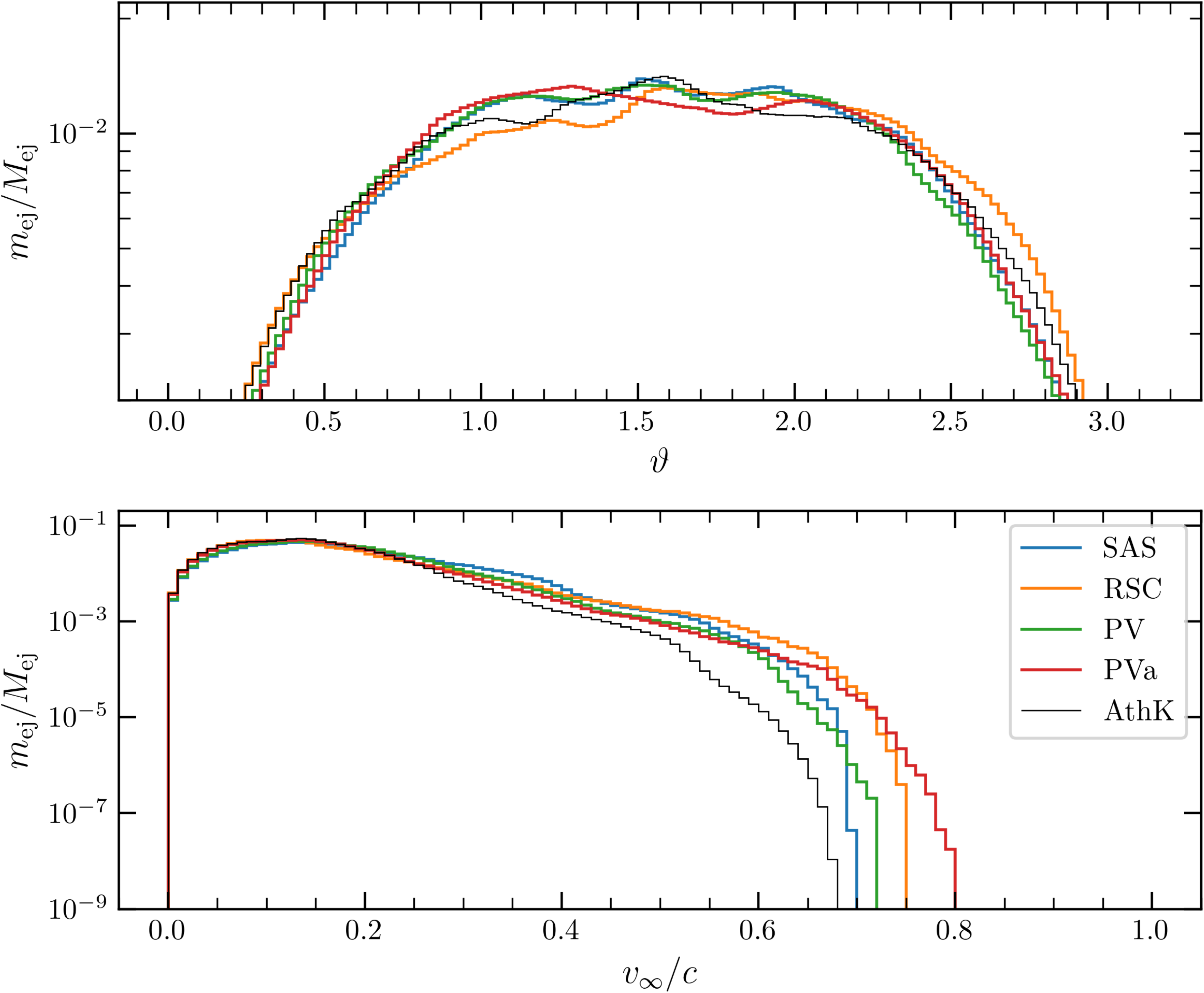}
  \caption{Mass-weighted histograms of ejecta angular distribution
    (top) and asymptotic velocity (bottom).
    Unbound mass is monitored according to the Bernoulli criterion and extracted at a
    sphere of coordinate radius $R\simeq 591\,\mbox{km}$. The cumulative
    histograms are computed at $t-t_{\rm mrgr}=21\,\mbox{ms}$. Common
    (finest) resolution of these runs in $\delta x_f$. \athenak{} data
    obtained with a very similar 
    simulation setup as \gra{} SAS are shown in black.}
 \label{fig:ejecta:histo}
\end{figure}

In order to verify that all atmosphere treatments reproduce the same
ejecta, we compute the mass-weighted histograms of the angular
distribution in the latitude angle $\vartheta$ with respect to the
orbital plane and of the asymptotic mass-averaged velocity. These quantities are
shown in Fig.~\ref{fig:ejecta:histo}. As expected for equal-mass
mergers, the bulk ejecta is concentrated around the orbital plane with
half opening angle of ${\sim}30$~degrees. The different vacuum
treatments in \gra{} do not significantly alter the spatial
distribution of the ejecta.

The asymptotic mass-averaged velocities also behave very similarly when
different vacuum treatments are considered (bottom panel of
Fig.~\ref{fig:ejecta:histo}.). Main differences are present in the
fast tail of the ejecta. The more aggressive pseudo-vacuum treatment
(PVa) can resolve fast tails with velocities up to ${\lesssim}0.8$c at
the considered resolution, while standard atmosphere treatment resolves
up to ${\sim0.65}$c in this simulation. Fast tails are composed of low density material
just above the atmosphere level; the launch and propagation of this
material is facilitated by the higher cutting value for the vacuum
threshold $D_{\rm cut}$ in PVa.

We remark that in the presented results the bulk of the ejecta is
mostly contained on the computational grid and start to outflow at
${\sim}20$~ms postmerger, see
Fig.~\ref{fig:cons}. For the simulated timescale, this does not
significantly affect the rescaling algorithm. However, for significantly longer evolution it
may be necessary to dynamically change the rescaling parameters
according to the physical mass outflowing the domain. This does not
pose major conceptual difficulties and could also be adopted for
matter accreting onto (through) an apparent horizon.

\subsection{Gravitational Waves}
\label{sbsbsec:gw}

\begin{figure}[t]
  \centering
  \includegraphics[width=0.49\textwidth]{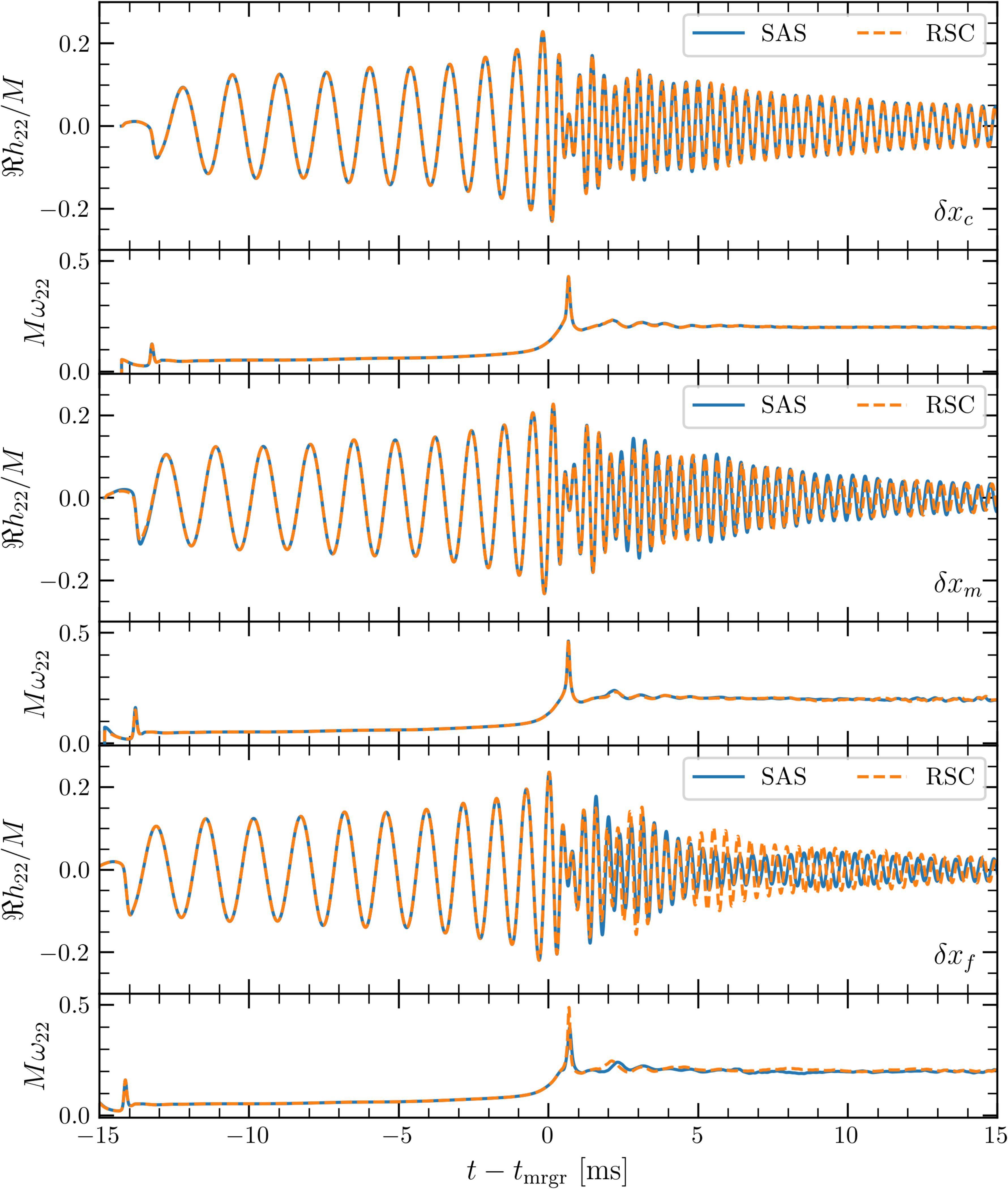}
  \caption{Gravitational waves strain and instantaneous frequency.
    Real part and instantaneous frequency of the $(\ell m)=(22)$ mode
    of the strain, based on evolution with the standard atmosphere
    treatment (SAS) in (blue, solid lines), and rescaling (RSC), in
    (orange, dashed lines).
    Top, middle, and bottom panels are computed from simulations
    involving finest resolutions of $\delta x_c$, $\delta x_m$, and
    $\delta x_f$ respectively. Waves are extracted at a sphere of
    coordinate radius $R\simeq295.3\,\mbox{km}$.}
  \label{fig:wvf}
\end{figure}

We next explore the computation of gravitational waves in
simulations with different atmosphere treatment.
Figure~\ref{fig:wvf} compares the dominant mode of the strain and
the instantaneous wave frequency computed with the standard atmosphere
and rescaling algorithm at different resolutions.
Minor differences in the wave amplitudes and phases are visible only at the
highest resolution and in the postmerger part of the wave. In
particular, the use of the rescaling algorithm appears to produce a
slightly more compact remnant towards the end of the
gravitational-wave dominated postmerger phase at these
resolutions. In turn, the waveform's amplitude is slightly larger
(less damped) at around $t\sim6$~ms postmerger. Similar results are
found for the pseudo-vacuum treatments.

\begin{figure}[t]
  \centering
  \includegraphics[width=0.49\textwidth]{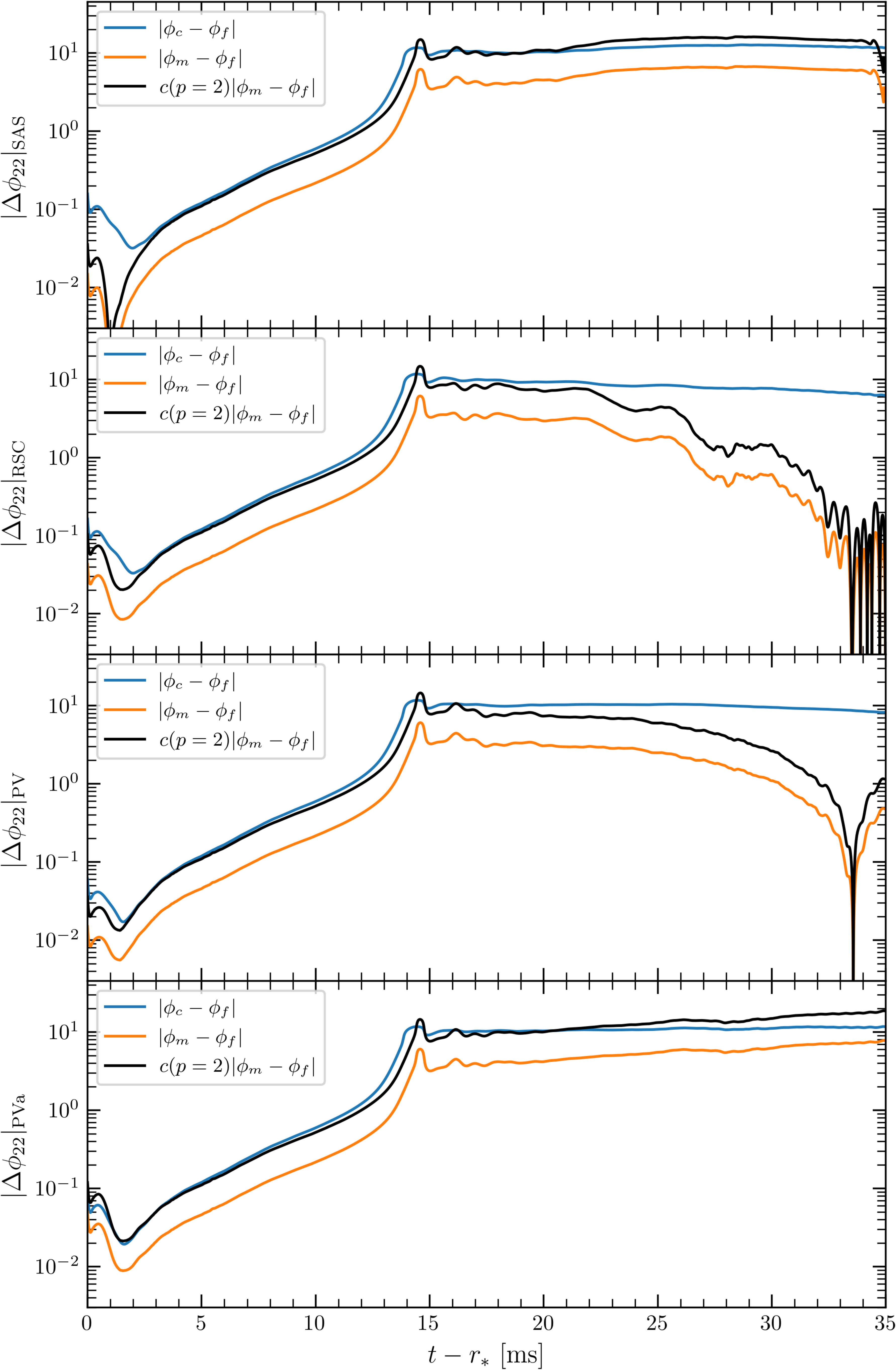}
  \caption{Self convergence of the phase of the gravitational wave's leading mode 
  extracted at $R\simeq295.3\,\mbox{km}$ for the different
  atmosphere treatments: SAS, RSC, PV and PVa from top to
  bottom respectively.}
  \label{fig:wvfconv}
\end{figure}

Waveform's phase self-convergence plots are shown in Fig.~\ref{fig:wvfconv}.
All the atmosphere treatments give consistent second-order convergent results 
in the inspiral phase of the evolution. In the postmerger phase, the standard
atmosphere treatment leads to second-order convergent waveforms, differently from
what previously observed in \cite{Daszuta:2024chz,Daszuta:2024ucu}. This
we attribute to LOFC, which was not previously available.
Waveform convergence progressively degrades in the postmerger phase for the
RSC and PV treatments at the considered resolution. The more
aggressive PVa treatment, however, does achieve approximate second-order
convergence. These results indicate that the combination of atmosphere
treatment and AMR flux-correction scheme are potentially key to achieve
postmerger waveform convergence already at the moderate resolutions
considered here.

\subsection{Comparison with \athenak}
\label{sbsbsec:comp}

We finally compare the vacuum treatments in \gra{} with
\athenak.
The same simulations presented above have been repeated using a very
similar, although not identical setup in the \athenak{} code.
The atmosphere treatment is very similar to the one in
\gra{}. However, \athenak{} conservative
AMR implements a different flux-correction: the FOFC described in~\cite{Lemaster:2008gh,Stone:2024,Fields:2024pob}.
The grid setup closely mirrors
the \gra{} setup, but the AMR criterion is slightly different
\cite{Fields:2024pob}. In particular, the resolution at the
wave (ejecta) extraction sphere of \athenak{} is
${\sim}3.94, 5.91, 3.94$~km (${\sim}7.88, 11.81, 7.88$~km) for the three
resolutions, while in \gra{} is ${\sim}7.88, 5.91, 7.88$~km
(${\sim}15.75, 11.81, 15.75$~km).
Outflows are computed using a latitude-longitude grid on the
extraction sphere with $N_\theta=256$, $N_\phi=512$ points and
coordinate radius $R\simeq591$ km.
Both codes extract GWs on geodesic grids at the extraction spheres
(\gra{} uses 9002 vertices and \athenak{} 1002) by computing the Weyl pseudoscalar
$\psi_4$ and integrating via fixed frequency integration (FFI)
\cite{Reisswig:2010di}.

The violation of baryon mass and lepton number conservation with the
standard atmosphere treatments in \gra{} and \athenak{} is shown in
Fig.~\ref{fig:cons}.  Despite implementation differences in the AMR and
flux-correction the performances of the two codes are very comparable.
While performing this comparison, we discovered an issue with scalar
conservation in \athenak{}, which produced violations at the $10^{-3}$
level. These emerged because $Y_e$ could become smaller than the minimum
value allowed by the EOS and would be floored. To improve mass
conservation, we have augmented the FOFC scheme with a positivity
preserving limiter, following the approach in \cite{Radice:2013xpa}.
Figure~\ref{fig:cons} shows the results obtained after having
implemented this fix.

Ejecta properties are contrasted in Fig.~\ref{fig:ejecta} and
Fig.~\ref{fig:ejecta:histo}.
\athenak{} simulation shows sustained ejecta velocity after ${\sim}10$ms
postmerger while, as mentioned above, \gra{} ejecta velocity goes more
rapidly to zero. This is due to the different structures of the AMR
Cartesian grid, which results in a coarser \gra{} grid towards and at
the extraction spheres of approximately a factor 2.
As the low density ejecta propagate outwards they are practically
absorbed into the atmosphere or the vacuum.
At the same time, the comparison confirms that the RSC and PV
treatments help to better resolve the fast tail of the bulk of dynamical
ejecta (see above).

\begin{figure}[t]
  \centering
  \includegraphics[width=0.49\textwidth]{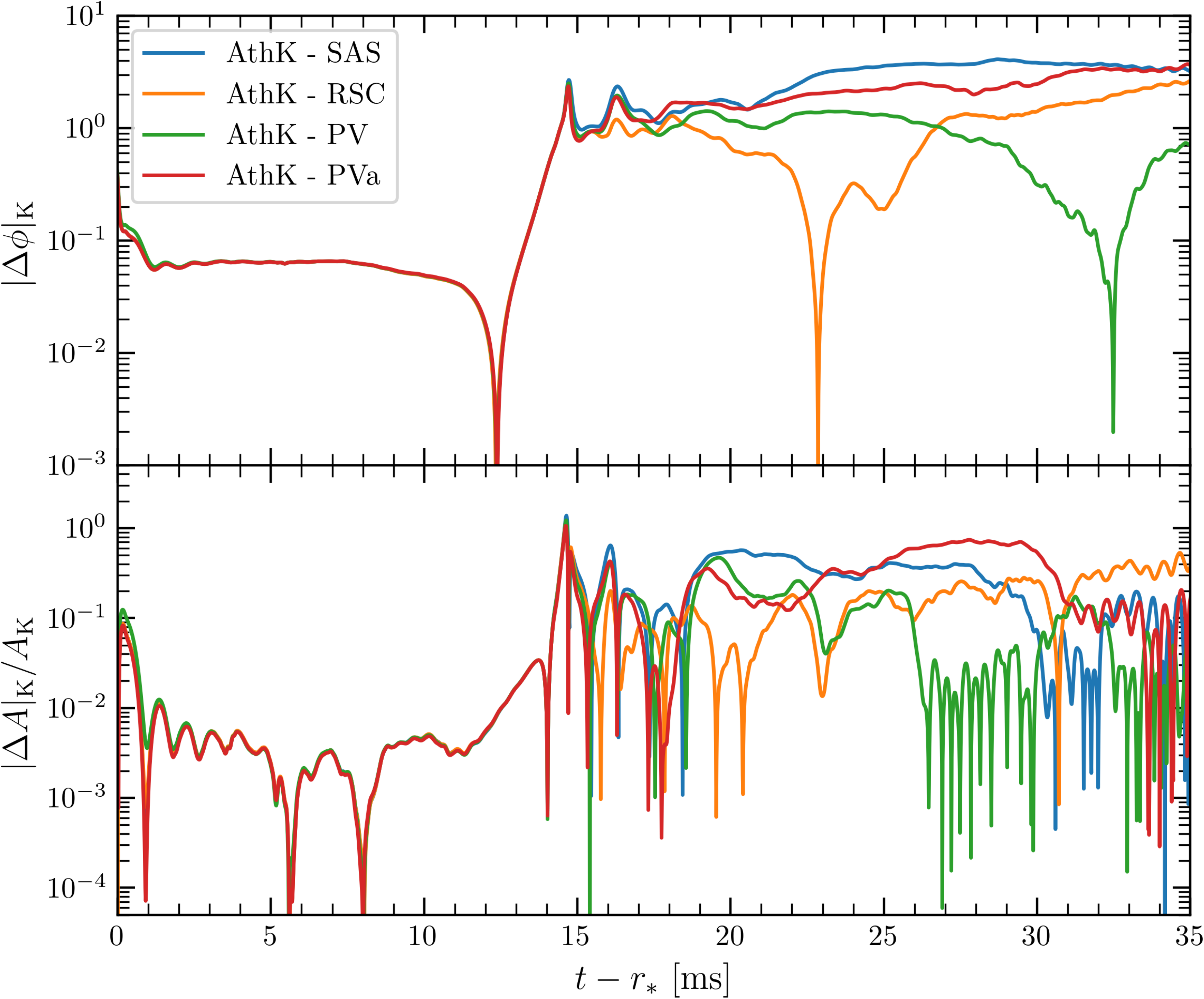}
  \caption{Gravitational waves phase (top) and amplitude (bottom) differences of 
  the four atmosphere treatments (SAS-blue, RSC-orange, PV-green and PVa-red)
  implemented in \gra{} with respect to \athenak{} data. Waves are
  extracted at a sphere of coordinate radius $R\simeq295.3\,\mbox{km}$.
  Simulations are at resolution $\delta x_f$.} 
  \label{fig:kwvfcomparison}
\end{figure}

Regarding the gravitational waveforms, 
Fig. \ref{fig:kwvfcomparison} shows the phase and amplitude (relative)
differences of the leading mode of the signal between \athenak{} 
and the four atmosphere treatments implemented in \gra{} at the highest considered
resolution $\delta x_f$. During the inspiral, the phase differences are consistently
lower than $0.1$rad (and flat) while amplitude relative differences are below $1\%$. 
These differences are comparable to (or smaller than) truncation
errors at the considered resolution.
As anticipated by Fig. \ref{fig:wvf}, differences are more prominent
in the postmerger phase. \athenak{} data are closer to the \gra{} RSC
and PV treatments. Larger differences are found with SAS and PVa at later
times, which however achieve convergence at the considered
resolutions (cf. Fig. \ref{fig:wvfconv}).
We note that, while the AMR criterion in \athenak{} guarantees higher
resolution at the extraction spheres than in \gra{}, it does not
generate a grid hierarchy that is suitable for a consistent
three-level convergence test in the wave zone.

\section{Conclusion}
\label{sec:conclusion}

In this paper, we proposed a simple yet efficient algorithm to ensure
exact mass conservation in Eulerian general-relativistic simulations
of neutron star spacetimes. In its simplest form, the scheme is a
dynamical rescaling of the low-density atmosphere floors that can be
easily incorporated in existing codes. Such a rescaling can be coupled
with a pseudo-vacuum treatment that enforces strict vacuum in exterior
regions and reduces simulation cost proportionally to the extent of
the latter.

We demonstrated the effectiveness of these vacuum treatments with
BNSM simulations spanning multiple orbits and
the GW-driven postmerger phase, and including a microphysical equation of state.
The comparison of gravitational waves and (dynamical) ejecta computed
with different atmosphere treatments indicates that these observables
are only very mildly affected by the vacuum treatment and are
therefore robust.
The comparison between \gra{} and \athenak{} clearly highlights that
implementation details matter. Moreover we found crucial the use of
conservative AMR with flux corrections for achieving robust
conservation of baryonic mass (and more broadly, other extant
conserved charges).

The proposed rescaling and pseudo-vacuum treatment may be relevant in
long-term evolutions of BNSM, also in the presence of neutrino
radiation. Improved mass conservation and pseudo-vacuum treatment can 
improve the accuracy of mass-outflow computation when typical
densities reach value comparable to the density floors. Such
simulations may require monitoring mass outflowing from the
computational domain and dynamically correcting the algorithmic
parameters,  ($M_b(0), f_D$) in Eq.~\eqref{eq:src:algo}.  
A similar adjustment could be required in presence of black hole
formation, in order to improve the remnant disc evolution.

The proposed scheme also allows exact electron number conservation in
absence of neutrino radiation. In presence of the latter, the scheme
should be adjusted taking into account the source terms in the
electron fraction balance law. Ultimately, the scheme may lead to an
improved lepton number conservation.
Analogously to the dynamical mass adjustment discussed above, there is
no immediate difficulty incorporating these changes but extensive
experimentation is likely needed.
These explorations are ongoing.

\begin{acknowledgments}
  BD, JF and SB acknowledge support by the EU Horizon under ERC Consolidator Grant,
  no. InspiReM-101043372. 
  DR acknowledges support from the U.S.~Department of Energy, Office of
  Science, Division of Nuclear Physics under Award Number(s)
  DE-SC0021177 and DE-SC0024388; and from the National Science Foundation under Grants
  No.~PHY-2020275, PHY-2116686, PHY-2407681, and PHY-2512802.
  DR and JMF acknowledge support from NASA under Award Number 80NSSC25K7213.

  \gra{} simulations were performed on the national HPE Apollo Hawk (Hunter)
  at the High Performance Computing Center Stuttgart (HLRS). 
  The authors acknowledge HLRS for funding this project by providing access 
  to the supercomputer HPE Apollo Hawk (Hunter) under the grant numbers INTRHYGUE/44215
  and MAGNETIST/44288. 
  Simulations were also performed on SuperMUC\_NG at the
  Leibniz-Rechenzentrum (LRZ) Munich.  
  The authors acknowledge the Gauss Centre for Supercomputing
  e.V. (\url{www.gauss-centre.eu}) for funding this project by providing
  computing time on the GCS Supercomputer SuperMUC-NG at LRZ
  (allocations {\tt pn67xo}, {\tt pn76li}, {\tt pn68wi} and {\tt pn36jo}).
  The authors gratefully acknowledge the computing time provided on the 
  high-performance computer Lichtenberg II at TU Darmstadt, funded by the German 
  Federal Ministry of Research, Technology and Space (BMFTR), and the State of 
  Hesse Ministry of Science and Research, Art and Culture (HMWK).
  \texttt{AthenaK} simulations were performed on Aurora at the Argonne
  Leadership Computing Facility, which is a DOE Office of Science User
  Facility supported under Contract DE-AC02-06CH11357. An award of
  computer time was provided by the INCITE program. \texttt{AthenaK}
  simulations were also performed on Perlmutter, at the National Energy
  Research Scientific Computing Center (NERSC), a Department of Energy
  User Facility, using NERSC award ERCAP0031370.
  Postprocessing and development runs were performed on the ARA cluster
  at Friedrich Schiller University Jena. 
  The ARA cluster is funded in part by DFG grants INST
  275/334-1 FUGG and INST 275/363-1 FUGG, and ERC Starting Grant, grant
  agreement no. BinGraSp-714626.
\end{acknowledgments}

\appendix

\end{document}